\documentclass[aps,twocolumn,showpacs]{revtex4}
\usepackage{amsmath}
\usepackage{epsfig}

\begin{document}

\title{Prediction of $p\bar{\Omega}$ states and femtoscopic study}

\author{Ye Yan$^1$}\email{221001005@njnu.edu.cn}
\author{Qi Huang$^1$}\email{06289@njnu.edu.cn}
\author{Qian Wu$^{2,3}$}\email{qwu@nju.edu.cn}
\author{Hongxia Huang$^1$}\email{hxhuang@njnu.edu.cn}
\author{Jialun Ping$^1$}\email{jlping@njnu.edu.cn}
\affiliation{$^1$Department of Physics, Nanjing Normal University, Nanjing 210023, China}
\affiliation{$^2$School of Physics, Nanjing University, Nanjing 21000, China}
\affiliation{$^3$Institute of Modern Physics, Chinese Academy of Sciences, Lanzhou 730000, China}

\begin{abstract}
Inspired by recent researches on the $p \Omega$ and $p \bar{\Lambda}$ systems, we investigate the $p \bar{\Omega}$ systems within the framework of a quark model.
Our results show that the attraction between a nucleon and $\bar{\Omega}$ is slightly stronger than that between a nucleon and $\Omega$, suggesting that the $p \bar{\Omega}$ system is more likely to form bound states.
The dynamic calculations indicate that the $p \bar{\Omega}$ systems with both $J^{P}=1^{-}$ and $2^{-}$ can form bound states, with binding energies deeper than those of the $p \Omega$ systems with $J^{P}=2^{+}$.
The scattering phase shift and scattering parameter calculations also support the existence of $p \bar{\Omega}$ states.
Additionally, we discuss the behavior of the femtoscopic correlation function for the $p \bar{\Omega}$ pairs for the first time.
Considering the significant progress in experimental measurements of the correlation function of the $p \Omega$ system, the further study of the $p \bar{\Omega}$ systems using femtoscopic techniques will be a very valuable work.
\end{abstract}


\maketitle

\setcounter{totalnumber}{5}

\section{Introduction}
\label{sec:introduction}

The study of baryon-antibaryon bound states dates back to the proposal by Fermi and Yang~\cite{Fermi:1949voc} to form the pion from a nucleon-antinucleon pair.
In the traditional one-boson-exchange theory of nucleon-nucleon interactions, it is shown that the nucleon-antinucleon system is more attractive than the nucleon-nucleon system due to the strong $\omega$-exchange~\cite{Erkelenz:1974uj}.
Therefore, possible bound states and resonances of nucleon-antinucleon systems have been proposed for many years.
An extensive and comprehensive review of the possible bound states of $N\bar{N}$ was provided in Ref.\cite{Richard:1999qh}.

More recently, the BESIII Collaboration reported the observation of a new $X(1880)$ state in the line shape of the $3(\pi^+ \pi^-)$ invariant mass spectrum\cite{BESIII:2023vvr}, which is considered as evidence for the existence of a proton-antiproton bound state.
Many theoretical works have been sparked to study the $p \bar{p}$ system and the properties of $X(1880)$~\cite{Salnikov:2023ipo,Xiao:2024jmu,Karliner:2024cql,Niu:2024cfn,Ortega:2024zjx,Jia:2024ybo,Yang:2024idy}.
In addition to the possible proton-antiproton state, there has also been great progress in recent work related to $p\bar{\Lambda}$~\cite{BESIII:2023kgz}.
A narrow structure in the $p\bar{\Lambda}$ system near the mass threshold, named $X(2085)$, is observed in the process $e^+e^- \to pK^-\bar{\Lambda}$ with a statistical significance exceeding $20\sigma$.
And its spin and parity are slightly favored to be $J^P = 1^+$ through an amplitude analysis.
Further theoretical results and discussions can be found in Refs.~\cite{Li:2024hrf,Haidenbauer:2024smo,Zhang:2024ulk}.
Building on the significant progress in the studies of nucleon-antinucleon and nucleon-$\bar{\Lambda}$ systems, it is natural to explore whether bound states or resonance states can be formed between nucleon and other hyperons, or between nucleon and other antihyperons.

In recent years, the progress in understanding the strange dibaryon $p \Omega$ has renewed interest in dibaryon systems.
The STAR Collaboration measured the $p \Omega$ correlation functions in Au+Au collisions at the Relativistic Heavy-Ion Collider (RHIC)\cite{STAR:2018uho} and reported a positive scattering length for the $p \Omega$ interaction, which supports the hypothesis of a $p \Omega$ bound state.
In addition, the ALICE Collaboration reported measurements of the $p$-$\Omega$ correlation in $p$ + $p$ collisions at $\sqrt{s} = 13$ TeV at the Large Hadron Collider (LHC)\cite{ALICE:2020mfd}.
Beyond the $p \Omega$ system, femtoscopic techniques and correlation function studies have made significant progress, both experimentally~\cite{STAR:2014dcy, STAR:2015kha, ALICE:2018nnl, ALICE:2019hdt, ALICE:2019gcn, Fabbietti:2020bfg, ALICE:2021cpv, ALICE:2021njx, ALICE:2022enj, STAR:2024lzt} and theoretically~\cite{Ohnishi:1998at,Morita:2014kza,Ohnishi:2016elb,Hatsuda:2017uxk,Haidenbauer:2018jvl,Kamiya:2019uiw,Haidenbauer:2020kwo,Ohnishi:2021ger,Ogata:2021mbo,Mrowczynski:2021bzy,Graczykowski:2021vki,Kamiya:2021hdb,Haidenbauer:2021zvr,Liu:2022nec,Liu:2023uly,Liu:2023wfo,Molina:2023oeu,Liu:2024nac,Vidana:2023olz,Sarti:2023wlg,Li:2024tof,Molina:2023jov,Feijoo:2024bvn,Albaladejo:2023wmv,Feijoo:2023sfe,Ikeno:2023ojl,Kamiya:2024diw,Albaladejo:2023pzq,Torres-Rincon:2023qll,Abreu:2024qqo,Li:2024tvo,Albaladejo:2024lam,Etminan:2024uvc,Jinno:2024rxw,Etminan:2024nak}.

The $S=-3$, $I=1/2$, $J=2$ $p \Omega$ state was predicted by Goldman \textit{et al.} as a narrow resonance in a relativistic quark model~\cite{Goldman:1987ma}.
M. Oka also proposed the existence of a quasi-bound state with $I(J^{P})=1/2(2^{+})$ using a constituent quark model~\cite{Oka:1988yq}.
A recent study by the HAL QCD Collaboration, using lattice QCD, reported that the $p \Omega$ state is indeed a bound state at a pion mass of $875$ MeV~\cite{HALQCD:2014okw}, and later, with nearly physical quark masses ($m_{\pi} \simeq 146$ MeV and $m_{K} \simeq 525$ MeV)\cite{HALQCD:2018qyu}.
Using the interactions obtained from ($2+1$)-flavor lattice QCD simulations, K. Morita \textit{et al.} studied the two-pair momentum correlation functions of the $p \Omega$ state in relativistic heavy-ion collisions to further investigate the existence of a $p \Omega$ bound state\cite{Morita:2016auo, Morita:2019rph}.
Additionally, this state has also been confirmed to be a bound state in the chiral quark model~\cite{Li:1999bc}.

By analogy to the nucleon-nucleon and nucleon-antinucleon systems, one might speculate that there is likely more attraction in the $p \bar{\Omega}$ channel than in the $p \Omega$ channel.
If the $p \Omega$ state can be confirmed through further experimental measurements, we hope to observe an even stronger signal for the $p \bar{\Omega}$ state in experiments.
Furthermore, the nucleon-antinucleon state would annihilate very quickly in the ground state due to the quark content of this system, making it challenging to provide a convincing theoretical confirmation of the nucleon-antinucleon bound state or resonance.
In contrast, the $p \bar{\Omega}$ state cannot annihilate into the vacuum, as nucleon consists of three $u$($d$) quarks and $\bar{\Omega}$ consists of three $\bar{s}$ quarks.
Therefore, the $p \bar{\Omega}$ state is more stable and special, providing an ideal system for studying baryon-antibaryon interactions.
The copious production of anti-baryons in high-energy colliders offers excellent opportunities to study this type of spectrum.
Clearly, the theoretical study of the $p \bar{\Omega}$ system is both interesting and necessary, as it can provide valuable insights for experimental searches of baryon-antibaryon bound states.

In our previous work, we studied the $p \Omega$ interactions and correlation functions based on the quark delocalization color screening model (QDCSM)\cite{Yan:2024aap}.
According to our calculations, the depletion of the $p \Omega$ correlation functions caused by the $J^P = 2^+$ bound state, which is not observed in the ALICE Collaboration's measurements~\cite{ALICE:2020mfd}, can be explained by the contribution of the attractive $J^P = 1^+$ component in spin-averaging.
The QDCSM is a constituent quark model~\cite{Wang:1992wi, Wu:1996fm} that introduces two key ingredients: first, quark delocalization, which accounts for orbital excitation by allowing quarks to delocalize from one cluster to another; second, the color screening factor, which modifies the confinement interaction between quarks in different cluster orbits.
In the study of nucleon-nucleon and nucleon-hyperon interactions and the properties of the deuteron, the mechanism of quark delocalization and color screening plays a crucial role in generating intermediate-range attraction~\cite{Ping:1998si, Wu:1998wu, Pang:2001xx}.
This model has also been used to investigate various dibaryon candidates, such as $d^{*}$\cite{Ping:2008tp}, $p \Omega$\cite{Yan:2024aap,Huang:2015yza}, and others~\cite{Yan:2021glh,Yan:2022nxp,Yan:2023kqf,Yan:2023tvl,Yan:2023iie}.
It has been extended to study baryon-antibaryon systems, including $p\bar{p}$ and $p\bar{\Lambda}$~\cite{Huang:2011zzf, Huang:2011zq}.
Extending it to the $p \bar{\Omega}$ system is a natural progression.
Therefore, we continue to investigate the $p \bar{\Omega}$ system within the framework of the QDCSM.
In this work, the $p \bar{\Omega}$ system is studied from three aspects: energy spectrum, scattering processes, and correlation functions.

This paper is organized as follows.
A brief introduction of the QDCSM is given in the next section.
The correlation function and the inverse scattering method are introduced in Section~\ref{iCF} and Section~\ref{GLM}, respectively.
Section~\ref{discussion} devotes to the numerical results and discussions.
The summary is shown in the last section.

\section{THEORETICAL FORMALISM}
\subsection{Quark delocalization color screening model}

The detail of QDCSM used in the present work can be found in the references~\cite{Wang:1992wi,Wu:1996fm,Ping:1998si,Wu:1998wu,Pang:2001xx}.
Here, we present the salient features of the model.
The model Hamiltonian is given by:
\begin{align}
	H = & \sum_{i=1}^6 \left(m_i+\frac{\boldsymbol{p}_{i}^{2}}{2m_i}\right) -T_{\mathrm{c} . \mathrm{m}} + \sum_{j>i=1}^3 V_{qq}(\boldsymbol{r}_{ij}) \nonumber \\
	& + \sum_{j>i=4}^6 V_{\bar{q}\bar{q}}(\boldsymbol{r}_{ij})+ \sum_{i=1}^{3} \sum_{j=4}^{6} V_{q\bar{q}}(\boldsymbol{r}_{ij}),
\label{H}
\end{align}
where $m_i$ is the quark mass, $\boldsymbol{p}_{i}$ is the momentum of the quark, and $T_{\mathrm{c.m.}}$ is the center-of-mass kinetic energy.
The dynamics of the hexaquark system is driven by two-body potentials, including color confinement ($V_{\mathrm{CON}}$), perturbative one-gluon exchange interaction ($V_{\mathrm{OGE}}$), and dynamical chiral symmetry breaking ($V_{\chi}$).
\begin{align}
	V(\boldsymbol{r}_{ij})= & V_{\mathrm{CON}}(\boldsymbol{r}_{ij})+V_{\mathrm{OGE}}(\boldsymbol{r}_{ij})+V_{\chi}(\boldsymbol{r}_{ij}).
\end{align}

Here, a phenomenological color screening confinement potential ($V_{\mathrm{CON}}$) is used as:
\begin{align}
	V_{\mathrm{CON}}(\boldsymbol{r}_{ij}) = & -a_{c}\boldsymbol{\lambda}_{i}^{c} \cdot \boldsymbol{\lambda}_{j}^{c}\left[  f(\boldsymbol{r}_{ij})+V_{0}\right],
\label{CON}
\end{align}
\begin{align}
	f(\boldsymbol{r}_{ij}) =& \left\{\begin{array}{l}
		\boldsymbol{r}_{i j}^{2}, ~~~~~~~~~~~~~ ~i,j ~\text {occur in the same cluster} \\
		\frac{1-e^{-\mu_{q_{i}q_{j}} \boldsymbol{r}_{i j}^{2}}}{\mu_{q_{i}q_{j}}},  ~~~i,j ~\text {occur in different cluster}
	\end{array}\right.   \nonumber
\end{align}
where $a_c$, $V_{0}$ and $\mu_{q_{i}q_{j}}$ are model parameters, and $\boldsymbol{\lambda}^{c}$ stands for the SU(3) color Gell-Mann matrices.
Among them, the color screening parameter $\mu_{q_{i}q_{j}}$ is determined by fitting the deuteron properties, nucleon-nucleon scattering phase shifts, and hyperon-nucleon scattering phase shifts, respectively, with $\mu_{qq}=0.45$, $\mu_{qs}=0.19$, and $\mu_{ss}=0.08~$fm$^{-2}$, satisfying the relation---$\mu_{qs}^{2}=\mu_{qq}\mu_{ss}$~\cite{Chen:2011zzb}.
The one-gluon exchange potential ($V_{\mathrm{OGE}}$) is written as:
\begin{align}
	V_{\mathrm{OGE}}(\boldsymbol{r}_{ij})= &\frac{1}{4}\alpha_{s_{q_i q_j}} \boldsymbol{\lambda}_{i}^{c} \cdot \boldsymbol{\lambda}_{j}^{c} \nonumber  \\
	&\cdot \left[\frac{1}{r_{i j}}-\frac{\pi}{2} \delta\left(\mathbf{r}_{i j}\right)\left(\frac{1}{m_{i}^{2}}+\frac{1}{m_{j}^{2}}+\frac{4 \boldsymbol{\sigma}_{i} \cdot \boldsymbol{\sigma}_{j}}{3 m_{i} m_{j}}\right)\right],
\label{OGE}
\end{align}
where $\boldsymbol{\sigma}$ is the Pauli matrices and $\alpha_{s}$ is the quark-gluon coupling constant.
In order to cover the wide energy range from light to strange quarks, an effective scale-dependent quark-gluon coupling $\alpha_{s}( \mu )$ was introduced \cite{Vijande:2004he}:
\begin{eqnarray}
	\alpha_{s}(\mu) & = &
	\frac{\alpha_{0}}{\ln(\frac{\mu^2+\mu_{0}^2}{\Lambda_{0}^2})}.
\end{eqnarray}
The dynamical breaking of chiral symmetry results in the SU(3) Goldstone boson exchange interactions appear between constituent light quarks $u, d$, and $s$.
Hence, the chiral interaction is expressed as
\begin{align}
	V_{\chi}(\boldsymbol{r}_{ij})= & V_{\pi}(\boldsymbol{r}_{ij})+V_{K}(\boldsymbol{r}_{ij})+V_{\eta}(\boldsymbol{r}_{ij}).
\end{align}
Among them,
\begin{align}
	V_{\pi}\left(\boldsymbol{r}_{i j}\right) =&\frac{g_{c h}^{2}}{4 \pi} \frac{m_{\pi}^{2}}{12 m_{i} m_{j}} \frac{\Lambda_{\pi}^{2}}{\Lambda_{\pi}^{2}-m_{\pi}^{2}} m_{\pi}\left[Y\left(m_{\pi} \boldsymbol{r}_{i j}\right)\right. \nonumber \\
	&\left.-\frac{\Lambda_{\pi}^{3}}{m_{\pi}^{3}} Y\left(\Lambda_{\pi} \boldsymbol{r}_{i j}\right)\right]\left(\boldsymbol{\sigma}_{i} \cdot \boldsymbol{\sigma}_{j}\right) \sum_{a=1}^{3}\left(\boldsymbol{\lambda}_{i}^{a} \cdot \boldsymbol{\lambda}_{j}^{a}\right),
\label{pi}
\end{align}
\begin{align}
	V_{K}\left(\boldsymbol{r}_{i j}\right) =&\frac{g_{c h}^{2}}{4 \pi} \frac{m_{K}^{2}}{12 m_{i} m_{j}} \frac{\Lambda_{K}^{2}}{\Lambda_{K}^{2}-m_{K}^{2}} m_{K}\left[Y\left(m_{K} \boldsymbol{r}_{i j}\right)\right. \nonumber \\
	&\left.-\frac{\Lambda_{K}^{3}}{m_{K}^{3}} Y\left(\Lambda_{K} \boldsymbol{r}_{i j}\right)\right]\left(\boldsymbol{\sigma}_{i} \cdot \boldsymbol{\sigma}_{j}\right) \sum_{a=4}^{7}\left(\boldsymbol{\lambda}_{i}^{a} \cdot \boldsymbol{\lambda}_{j}^{a}\right),
\label{K}
\end{align}
\begin{align}
	V_{\eta}\left(\boldsymbol{r}_{i j}\right) =&\frac{g_{c h}^{2}}{4 \pi} \frac{m_{\eta}^{2}}{12 m_{i} m_{j}} \frac{\Lambda_{\eta}^{2}}{\Lambda_{\eta}^{2}-m_{\eta}^{2}} m_{\eta}\left[Y\left(m_{\eta} \boldsymbol{r}_{i j}\right)\right. \nonumber \\
	&\left.-\frac{\Lambda_{\eta}^{3}}{m_{\eta}^{3}} Y\left(\Lambda_{\eta} \boldsymbol{r}_{i j}\right)\right]\left(\boldsymbol{\sigma}_{i} \cdot \boldsymbol{\sigma}_{j}\right)\left[\cos \theta_{p}\left(\boldsymbol{\lambda}_{i}^{8} \cdot \boldsymbol{\lambda}_{j}^{8}\right)\right.  \nonumber \\
	&\left.-\sin \theta_{p}\left(\boldsymbol{\lambda}_{i}^{0} \cdot \boldsymbol{\lambda}_{j}^{0}\right)\right],
\label{eta}
\end{align}
where $Y(x) = e^{-x}/x$ is the standard Yukawa function.
The physical $\eta$ meson is considered by introducing the angle $\theta_{p}$ instead of the octet one.
The $\boldsymbol{\lambda}^a$ are the SU(3) flavor Gell-Mann matrices.
The values of $m_\pi$, $m_k$ and $m_\eta$ are the masses of the SU(3) Goldstone bosons, which adopt the experimental values~\cite{ParticleDataGroup:2024cfk}.
The chiral coupling constant $g_{ch}$, is determined from the $\pi N N$ coupling constant through
\begin{align}
	\frac{g_{c h}^{2}}{4 \pi} & = \left(\frac{3}{5}\right)^{2} \frac{g_{\pi N N}^{2}}{4 \pi} \frac{m_{u, d}^{2}}{m_{N}^{2}}.
\end{align}
Assuming that flavor SU(3) is an exact symmetry, it will only be broken by the different mass of the strange quark.
The other symbols in the above expressions have their usual meanings.

As for $V_{\bar{q}\bar{q}}(\boldsymbol{r}_{ij})$ and $V_{q\bar{q}}(\boldsymbol{r}_{ij})$ in Eq.~(\ref{H}), which represent the antiquark-antiquark ($\bar{q}\bar{q}$) and quark-antiquark ($q\bar{q}$) interactions.
For the antiquark, replacing $\boldsymbol{\lambda}_i$ in Eqs.~(\ref{CON}) and (\ref{OGE}) with $-\boldsymbol{\lambda}^{*}_i$, and $\boldsymbol{\lambda}^a_i$ in Eqs.~(\ref{pi}),~(\ref{K}), and (\ref{eta}) with $\boldsymbol{\lambda}^{a*}_i$, the forms of $V_{\bar{q}\bar{q}}$ and $V_{q\bar{q}}$ can be derived.
It is noted that there is no annihilation between quark and antiquark.
The reason is that the $p \bar{\Omega}$ state cannot annihilate to the vacuum due to the different quark flavor contents of $N$ and $\bar{\Omega}$.
All the parameters are taken from our previous work of the $p \Omega$ systems~\cite{Huang:2015yza}, fixed by masses of the ground-state baryons.

In addition, quark delocalization was introduced to enlarge the model variational space to take into account the mutual distortion or the internal excitations of nucleons in the course of interaction.
It is realized by specifying the single-particle orbital wave function of the QDCSM as a linear combination of left and right Gaussians, the single-particle orbital wave functions used in the ordinary quark cluster model
\begin{eqnarray}
	\psi_{\alpha}(\boldsymbol {S_{i}} ,\epsilon) & = & \left(
	\phi_{\alpha}(\boldsymbol {S_{i}})
	+ \epsilon \phi_{\alpha}(-\boldsymbol {S_{i}})\right) /N(\epsilon), \nonumber \\
	\psi_{\beta}(-\boldsymbol {S_{i}} ,\epsilon) & = &
	\left(\phi_{\beta}(-\boldsymbol {S_{i}})
	+ \epsilon \phi_{\beta}(\boldsymbol {S_{i}})\right) /N(\epsilon), \nonumber \\
	N(S_{i},\epsilon) & = & \sqrt{1+\epsilon^2+2\epsilon e^{-S_i^2/4b^2}}. \label{1q}
\end{eqnarray}
It is worth noting that the mixing parameter $\epsilon$ is not an adjusted one but determined variationally by the dynamics of the multiquark system itself.
In this way, the multiquark system chooses its favorable configuration in the interacting process.
This mechanism has been used to explain the crossover transition between the hadron phase and quark-gluon plasma phase~\cite{Xu:2007oam}.

\subsection{Two-particle correlation function}
\label{iCF}

Experimentally, the correlation function $C(\boldsymbol{k})$ can be measured based on:
\begin{align}
	C(\boldsymbol{k}) & = \xi(\boldsymbol{k}) \frac{N_{\text{same}}(\boldsymbol{k})}{N_{\text{mixed}}(\boldsymbol{k})},
\end{align}
where $N_{\text{same}}(\boldsymbol{k})$ and $N_{\text{mixed}}(\boldsymbol{k})$ represent the $\boldsymbol{k}$ distributions of hadron-hadron pairs produced in the same
and in different collisions, respectively, and $\xi(\boldsymbol{k})$ denotes the corrections for experimental effects.
In theoretical studies, the correlation function can be calculated using the Koonin$-$Pratt (KP) formula~\cite{Koonin:1977fh,Pratt:1990zq,Bauer:1992ffu}:
\begin{align}
	C(\boldsymbol{k}) & = \frac{N_{12}\left(\boldsymbol{p}_{1}, \boldsymbol{p}_{2}\right)}{N_{1}\left(\boldsymbol{p}_{1}\right) N_{2}\left(\boldsymbol{p}_{2}\right)} \nonumber \\
	& \simeq \frac{\int \mathrm{d}^{4} x_{1} \mathrm{~d}^{4} x_{2} S_{1}\left(x_{1}, \boldsymbol{p}_{1}\right) S_{2}\left(x_{2}, \boldsymbol{p}_{2}\right)|\Psi(\boldsymbol{r}, \boldsymbol{k})|^{2}}{\int \mathrm{d}^{4} x_{1} \mathrm{~d}^{4} x_{2} S_{1}\left(x_{1}, \boldsymbol{p}_{1}\right) S_{2}\left(x_{2}, \boldsymbol{p}_{2}\right)} \nonumber \\
	& \simeq \int \mathrm{d} \boldsymbol{r} S_{12}(r)|\Psi(\boldsymbol{r}, \boldsymbol{k})|^{2},
\label{ignore}
\end{align}
where $S_{i}(x_{i}, \boldsymbol{p}_{i})~(i = 1, 2)$ is the single particle source function of the hadron $i$ with momentum $\boldsymbol{p}_{i}$, $\boldsymbol{k} = (m_2 \boldsymbol{p}_{1} - m_1 \boldsymbol{p}_{2})/(m_1 + m_2)$ is the relative momentum in the center-of-mass of the pair $(\boldsymbol{p}_{1} + \boldsymbol{p}_{2} = 0)$, $\boldsymbol{r}$ is the relative coordinate with time difference correction, and $\Psi(\boldsymbol{r}, \boldsymbol{k})$ is the relative wave function in the two-body outgoing state with an asymptotic relative momentum $\boldsymbol{k}$.
In the case where we can ignore the time difference of the emission and the momentum dependence of the source, we integrate out the center-of-mass coordinate and obtain Eq.~(\ref{ignore}), where $S_{12}(r)$ is the normalized pair source function in the relative coordinate,given by the expression:
\begin{align}
	S_{12}(r) = \frac{1}{(4 \pi R^2)^{3/2}} \text{exp}(-\frac{r^2}{4R^2}),
\label{source}
\end{align}
where $R$ is the size parameter of the source.
Thus, two important factors of the correlation function are included in Eq.~(\ref{ignore}): the collision system, which is related to the source function $S_{12}(r)$, and the two-particle interaction, which is embedded in the relative wave function $\Psi(\boldsymbol{r}, \boldsymbol{k})$.

For a pair of non-identical particles, such as $p \bar{\Omega}$, assuming that only $S$-wave part of the wave function is modified by the two-particle interaction, $\Psi(\boldsymbol{r}, \boldsymbol{k})$ can be given by:
\begin{align}
	\Psi_{p \bar{\Omega}}(\boldsymbol{r}, \boldsymbol{k}) = \text{exp}(\text{i} \boldsymbol{k} \cdot \boldsymbol{r}) -j_0(kr) + \psi_{p \bar{\Omega}}(r,k),
\end{align}
where the spherical Bessel function $j_0(kr)$ represents the $S$-wave part of the non-interacting wave function, and $\psi_{p \bar{\Omega}}$ stands for the scattering wave function affected by the two-particle interaction.
Substituting the relative wave function $\Psi_{p \bar{\Omega}}(\boldsymbol{r}, \boldsymbol{k})$ into the KP formula yields the correlation function:
\begin{align}
	C_{p \bar{\Omega}}(k) = 1 + \int_{0}^{\infty} 4\pi r^2 \, \mathrm{d}r \, S_{12}(r) \, [ |\psi_{p \bar{\Omega}}(r,k)|^2 - |j_0(kr)|^2 ]. \label{Ck}
\end{align}

$\psi_{p \bar{\Omega}}(r,k)$ can be obtained by solving the Schr\"{o}dinger equation, and a similar approach has been utilized in the femtoscopic correlation analysis tool using the Schr\"{o}dinger equation~\cite{Mihaylov:2018rva}:
\begin{align}
	-\frac{\hbar ^{2}}{2\mu }\nabla ^{2}\psi_{p \bar{\Omega}}(r,k) +V(r) \psi_{p \bar{\Omega}}(r,k) = E \psi_{p \bar{\Omega}}(r,k)
\end{align}
where $\mu = m_p m_\Omega / (m_p + m_\Omega)$ is the reduced mass of the system.

Considering the case of the $S$-wave, the wave function can be separated into a radial term $R_k(r)$ and an angular term $Y_0^0(\theta, \phi)$ and expressed as:
\begin{align}
	\psi_{p \bar{\Omega}}(r,\theta, \phi) =  R_k(r) Y_0^0(\theta, \phi).
\end{align}

Considering the interaction between a proton and an $\bar{\Omega}$ baryon, which includes both the strong interaction and the repulsive Coulomb interaction, the potential can be written as:
\begin{align}
	V(r) = V_{\text{Strong} }(r) + V_{\text{Coulomb}} (r),
\label{Coulomb}
\end{align}
where $V_{\text{Coulomb}}(r) = + \alpha \hbar c/r $, and $\alpha$ is the fine-structure constant.
The method to obtain the strong interaction potential $V_{\text{Strong} }(r)$ will be introduced in the next section.

Once the total interaction potential is determined, the radial Schr\"{o}dinger equation can be solved:
\begin{align}
	\frac{-\hbar^{2}}{2 \mu} \frac{\text{d}^{2}u_k(r)}{\text{d}r^{2}} + V(r) u_k(r)  = E u_k(r),   \label{eq}
\end{align}
where $E = \hbar^2 k^{2} / (2 \mu)$ and $u_k(r) = r R_k(r)$.
On this basis, the correlation function $C_{p \bar{\Omega}}(k)$ for given spin-parity quantum numbers can be calculated through Eq.~(\ref{Ck}).
The calculation of the correlation functions described above is based on obtaining the scattering wave functions by solving the Schr\"{o}dinger equation in coordinate space~\cite{Morita:2014kza,Ohnishi:2016elb,Hatsuda:2017uxk,Kamiya:2019uiw,Kamiya:2021hdb,Ohnishi:2021ger,Ogata:2021mbo}.
Additionally, the scattering wave functions can also be obtained by solving the Lippmann-Schwinger (Bethe-Salpeter) equation in momentum space~\cite{Haidenbauer:2018jvl,Haidenbauer:2020kwo,Liu:2022nec,Liu:2023uly,Liu:2024nac}.
Further details on correlation functions for various systems can be found in the references mentioned above.

Additionally, for the $S$-wave $p \bar{\Omega}$ dibaryon system, the possible spin-parity quantum numbers can be $J^P = 1^+$ and $2^+$, respectively.
Since the experimentally measured correlation function is spin-averaged, the theoretically obtained correlation function should also consider the average over systems with different quantum numbers:
\begin{align}
	C_{p \bar{\Omega}}(k) = \frac{3}{8}C_{p \bar{\Omega}}^{J = 1}(k) + \frac{5}{8} C_{p \bar{\Omega}}^{J = 2}(k).  \label{average}
\end{align}

\subsection{Gel'fand-Levitan-Marchenko method for inverse scattering problem}
\label{GLM}

Obviously, to solve Eq.~(\ref{eq}), two-body interaction potential $V(r)$ is absolutely necessary.
The QDCSM is actually a treatment on few-body problem, which means directly extracting a two-body interaction potential $V(r)$ from it will not be so natural since the hadronization process has not fully complete.
Fortunately, the QDCSM can be employed to investigate scattering process, which means we can use it to get the potential we need due to the completely finished hadronization there.

The approach we adopted to extract the two-body equivalent potential $V(r)$ is the GLM method, which is a very powerful tool in inverse scattering theory~\cite{Chadan1977}.
It can provide us a systematic approach to reconstruct an equivalent potential from the scattering data of a specific process, which makes it as a very classical ``inverse problem''.
Thus, this method will give us another path to understand the nature of two-body interaction.

The key equation of the GLM method used in the work is the Marchenko equation~\cite{Marchenko1955,Agranovich1963}, which can be written in the $S$-wave case in a integration equation form as:
\begin{align}
	K(r, r^{\prime}) + F(r, r^{\prime}) + \int_r^\infty K(r, s) F(s , r^{\prime}) \, \mathrm{d}s = 0.
\end{align}
Here, the kernel function $K(r, r^{\prime})$ is the solution of the equation to be determined, and $F(r, r^{\prime})$ is the inverse Fourier transformation of reflection coefficient as:
\begin{align}
	F(r, r^{\prime}) =& \frac{1}{2 \pi} \int_{-\infty}^{\infty} e^{\mathrm{i}kr}\left\{1-S(k)\right\} e^{\mathrm{i}kr^\prime} \mathrm{d} k \nonumber \\
	&+\sum_{i=1}^{n} M_{i} e^{-\kappa_i r} e^{-\kappa_i r^\prime}.
\end{align}
The partial-wave scattering matrix $S(k)$ is given by $S(k) = \mathrm{exp}(2 \mathrm{i} \delta(k))$, where $\delta(k)$ is the scattering phase shift satisfying $k \cot \delta = -1/a_0 + 1/2 \, r_{\text{eff}} k^2$.
Here, $a_0$ and $r_{\text{eff}}$ represent the scattering length and the effective range, respectively.
Additionally, $n$ is the number of bound states, $\kappa_i$ denotes the wavenumber of the $i$-th bound state, and $M_i$ is the norming constant.
Then, after solving Marchenko equation and obtaining $K(r,r^\prime)$, the potential can be reconstructed as:
\begin{equation}
	V(r)=-2\frac{\mathrm{d}}{\mathrm{d}r}K(r,r).
\end{equation}

There is one point we want to emphasize here.
Generally, when there exists bound states, this method can not give us a fully determined potential but end up with a set of phase-equivalent potentials~\cite{Sofianos:1990rb}.
However, if one fix all the $M_i$ in a unique way such as calculating from Jost solution, the obtained potential will be unique for further calculation~\cite{Massen:1999vi,Newton2002}.
By using this method, preparation for further calculation can be done, for a more comprehensive discussion on this method, one can refer to Refs.~\cite{Chadan1977,Marchenko1955,Agranovich1963,Sofianos:1990rb,Massen:1999vi,Newton2002,Jade:1996am,Meoto:2019jky,Khokhlov:2021afy,Khokhlov:2022uie}.

\section{RESULTS AND DISCUSSION}
\label{discussion}

The $S$-wave $p \bar{\Omega}$ systems with isospin $I=1/2$, spin parity $J^{P}=1^{-}$ and $2^{-}$ are investigated on the basis of the QDCSM.
In order to see whether or not there is any bound state, a dynamic calculation is performed as a first step.
Here we employ the resonating group method (RGM) to solve the bound-state problem, which is briefly introduced in the Appendix, and more details can be found in Refs.~\cite{Wheeler:1937zza,Kamimura:1977okl,Yan:2024usf}.
Expanding the relative motion wave function between two hadrons in the RGM equation by gaussians, the integro-differential equation of RGM can be reduced to an algebraic equation, the generalized eigen-equation.
Then the energy of the system can be obtained by solving the eigen-equation.
The binding energies of the $p \bar{\Omega}$ systems with $J^{P} = 1^{-}$ and $J^{P} = 2^{-}$, denoted as $E_B$, are listed in Table~\ref{bound}.
Here, $E^{\text{Theo}}_{\text{th}}$ represents the theoretical threshold, and $E^{\text{Theo}}$ represents the eigenvalue of the corresponding system.
The calculation for the $p \bar{\Omega}$ systems does not involve channel coupling, because we limit our study to color-singlet sub-clusters consisting of three $u/d$ quarks and three $\bar{s}$ quarks.

\begin{table}
\caption{The binding energies of the $p \bar{\Omega}$ systems (in MeV).}
\begin{tabular}{c c c c}
\hline \hline
 ~~~~$J^{P}$~~~~ & ~~~~$E^{\text{Theo}}_{\text{th}}$~~~~  & ~~~~$E^{\text{Theo}}$~~~~& ~~~~$E_B$~~~~  \\ \hline
    $~ 1^{-}$  & 2581 & 2571 & $10$  \\
    $~ 2^{-}$  & 2581 & 2572 & $9$   \\
\hline \hline
\label{bound}
\end{tabular}
\end{table}

From Table~\ref{bound} we can see that both the $J^{P}=1^{-}$ and $J^{P}=2^{-}$ $p \bar{\Omega}$ forms bound states with the binding energies about 10 MeV and 9 MeV, respectively.
By contrast, in our previous work on the $p \Omega$ systems, the single channel calculation shows that neither the $J^{P}=1^{+}$ nor $J^{P}=2^{+}$ $p \Omega$ is bound.
After channel coupling, only the $p \Omega$ with $J^{P}=2^{+}$ forms a bound state with binding energy about 6 MeV.
These numerical results indicate that it is more possible for the $p \bar{\Omega}$ system rather than the $p \Omega$ system to form bound states in our calculations.
Therefore, considering that the attractive $p \Omega$ interaction is implied in the experimental measurements of $p \Omega$ correlation functions~\cite{ALICE:2020mfd}, we look forward to the experimental progress on $p \bar{\Omega}$ correlation functions in future.

\begin{figure}[htb]
	\centering
	\includegraphics[width=6cm]{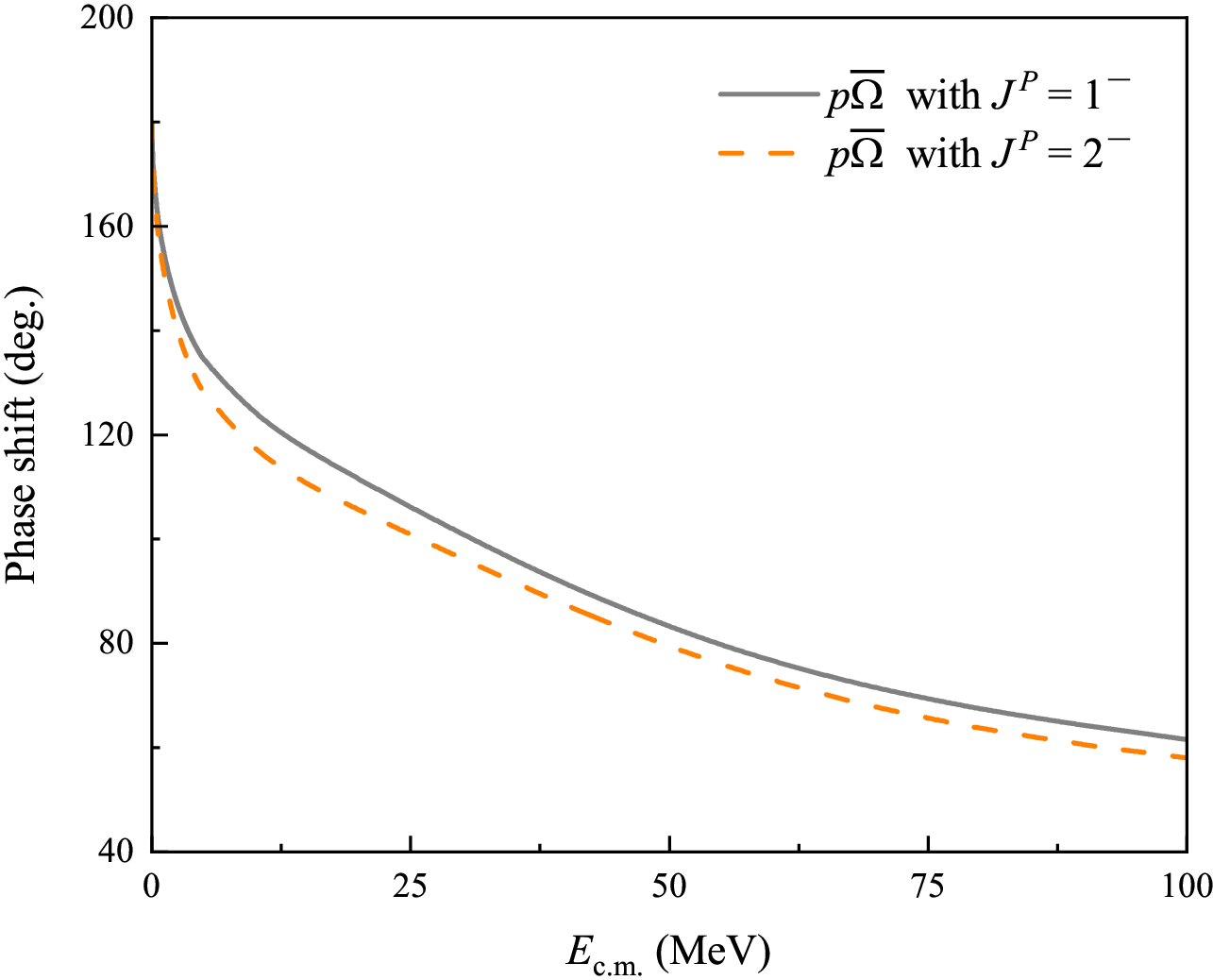}\
	\caption{The phase shifts of the $p \bar{\Omega}$ systems.}
	\label{shift}
\end{figure}

In order to further study the interaction between nucleon and $\bar{\Omega}$, we calculated the scattering phase shifts of the $p \bar{\Omega}$ systems.
The calculation is based on the well developed Kohn-Hulthen-Kato(KHK) variational method, the details of which can be found in Ref.~\cite{Kamimura:1977okl}.
We also give a brief introduction in the Appendix.
The low-energy scattering phase shifts of the $p \bar{\Omega}$ systems are shown in Fig.~\ref{shift}.
It should be noted that the strong interaction between nucleon and $\bar{\Omega}$ is the focus of both the bound state calculation discussed earlier and the scattering phase shift calculation, and therefore the Coulomb interaction is not considered.
The Coulomb interaction will be introduced in the next step when calculating the correlation functions.

As one can see, for the $p \bar{\Omega}$ systems with both $J^{P}=1^{-}$ and $J^{P}=2^{-}$, the scattering phase shifts go to $180$ degrees at $E_{\text{c.m.}} \rightarrow 0$ MeV and rapidly decreases as $E_{\text{c.m.}}$ increases, which indicates the existence of a bound state in these systems.
This conclusion is consistent with the the bound state calculation discussed earlier.
Then, we can extract the scattering length $a_{0}$ and the effective range $r_{\text{eff}}$ of the $p \bar{\Omega}$ systems from
the low-energy phase shifts obtained above by using the expansion:
\begin{eqnarray}
k\, cot\delta & = & -\frac{1}{a_{0}}+\frac{1}{2}r_{\text{eff}}k^{2}+{\cal O}(k^{4}),
\end{eqnarray}
where $k$ is the momentum of the relative motion with $k=\sqrt{2\mu E_{\mbox{c.m.}}}$, $\mu$ is the reduced mass of two baryons, and $E_{\mbox{c.m.}}$ is the incident energy; $\delta$ is the low-energy scattering phase shifts.
And the binding energy $E_{B}^{\prime}$ can be calculated according to the following relation:
\begin{eqnarray}
E_{B}^{\prime} & = &\frac{\hbar^2\alpha^2}{2\mu},
\label{E prime}
\end{eqnarray}
where $\alpha$ is the wave number which can be obtained from the relation~\cite{Babenko:2003js}:
\begin{eqnarray}
r_{\text{eff}} & = &\frac{2}{\alpha}(1-\frac{1}{\alpha a_{0}}).
\end{eqnarray}
Note that this is another way to calculate the binding energy, therefore it is labeled $E_{B}^{\prime}$.
The scattering parameters of the $p \bar{\Omega}$ systems, along with the binding energies obtained using the scattering parameters, are listed in Table~\ref{length}.

\begin{table}[htb]
\caption{The scattering length $a_{0}$, effective range $r_{\text{eff}}$, and binding energy $E_{B}^{\prime}$ of the $p \bar{\Omega}$ systems.}
\begin{tabular}{c c c c}
\hline \hline
  ~~$J^{P}$~~  & ~~$a_{0}$ (fm)~~ & ~~$r_{\text{eff}}$ (fm)~~ & ~~$E_{B}^{\prime}$ (MeV)~~    \\ \hline
    $ 1^{-}$ &  2.43 & 0.48 & 7  \\
    $ 2^{-}$ &  2.79 & 0.81 & 6  \\
\hline \hline
\label{length}
\end{tabular}
\end{table}

\begin{figure*}[htb]
	\centering
	\includegraphics[width=16cm]{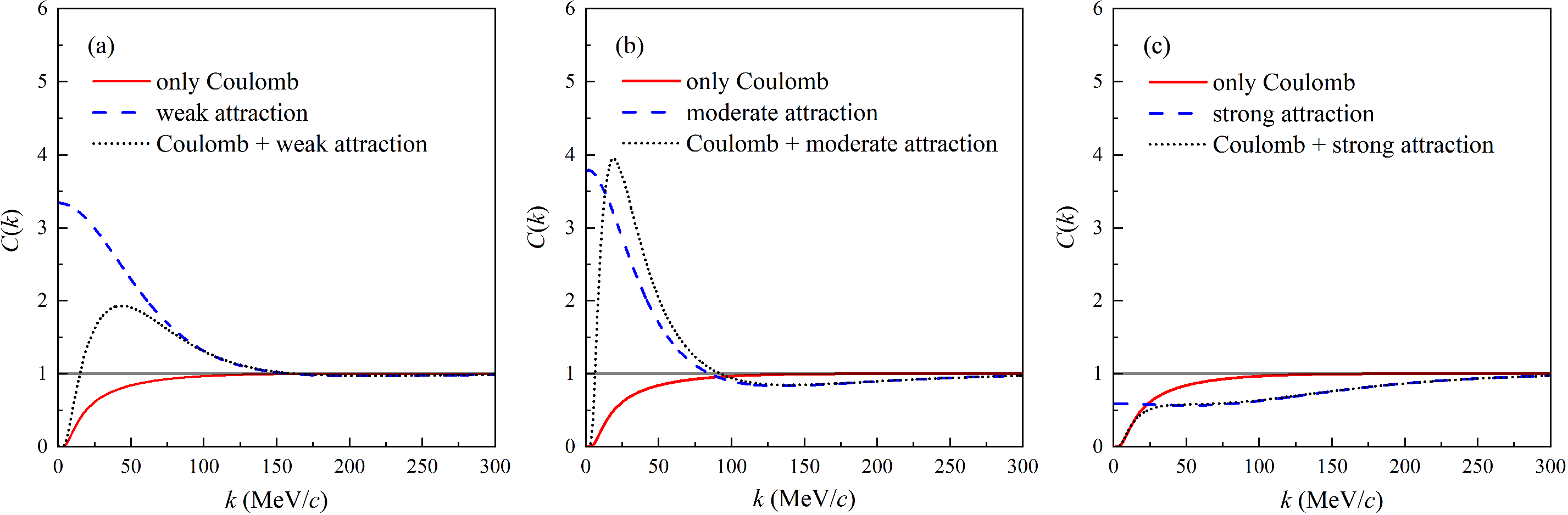}\
	\caption{The correlation functions for different square well potentials, with $V(r) = V_0 \, \theta(r_0 - r)$, where $V_0 = -10$ MeV in panel (a), $V_0 = -28$ MeV in panel (b), $V_0 = -40$ MeV in panel (c), and $r_0 = 2$ fm.}
	\label{well}
\end{figure*}

From Table~\ref{length}, we can see that the scattering lengths are positive for the $p \bar{\Omega}$ systems with $J^P = 1^-$ and $2^-$, which also confirms the existence of bound states.
Besides, the binding energies of the two systems obtained by Eq.~(\ref{E prime}) is broadly consistent with the numerical results shown in Table~\ref{bound}, which is obtained by the dynamic calculation.
Additionally, in the method of obtaining the binding energies using scattering parameters, the binding energies of the two $p \bar{\Omega}$ systems are also  slightly deeper than that of the $p \Omega$ system with $J^P= 2^+$.

Moreover, by solving the inverse scattering problem, we can further study the behavior of the $p \bar{\Omega}$ correlation functions on the basis of the $p \bar{\Omega}$ scattering process and the KP formula in Eq.~(\ref{ignore}).
Before that, we can study the general properties of the $p \bar{\Omega}$ correlation functions through an effective square well potential model.
The correlation functions corresponding to different degrees of square well potentials are presented in Fig.~\ref{well}.
The solid red lines represent the correlation functions that only influenced by the repulsive Coulomb interaction.
The dashed blue lines represent the correlation functions that only influenced by squire well potentials.
We introduce three square well potentials with width $r_0 = 2$ fm, weak attraction in panel(a) corresponding to $V_0 = -10$ MeV, moderate attraction in panel (b) corresponding to $V_0 = -28$ MeV, and relatively strong attraction in panel (c) corresponding to $V_0 = -40$ MeV.
The dotted black lines represent the correlation functions that influenced by both Coulomb interaction and squire well potentials through Eq.~(\ref{Coulomb}).
Additionally, the value of source size parameter $R$ in Eq.~(\ref{source}) are taken from our previous work on the $p \Omega$ correlation functions.

In Fig.~\ref{well}, panels (a) and (b), the correlation function affected only by the square potential is above unity in the low-energy region, which is due to the attractive interaction.
The difference is that the weak attraction in panel (a) is not enough to form a bound state, therefore the correlation function is always above unity, while the moderate attraction in panel (b) forms a shallow bound state.
The existence of the bound state leads to the depletion of the correlation function, so there exists a part below unity.
After taking into account both the Coulomb interaction and squire well potentials, which mainly dominates the low-energy region (0$<k<$25 MeV), the correlation function forms a peak-like structure.
In panel (c), the correlation function remains below unity for a relatively strong attraction.
A discussion about this phenomenon can be found in Refs.~\cite{Liu:2023uly,Yan:2024aap}.
After considering the Coulomb interaction, one can see that the correlation function in the low-energy region nearly coincides with the result obtained by considering only the Coulomb interaction.
As the relative momentum $k$ increases, the correlation function closely matches the result obtained by considering only a relatively strong attraction.

\begin{figure*}[htb]
	\centering
	\includegraphics[width=16cm]{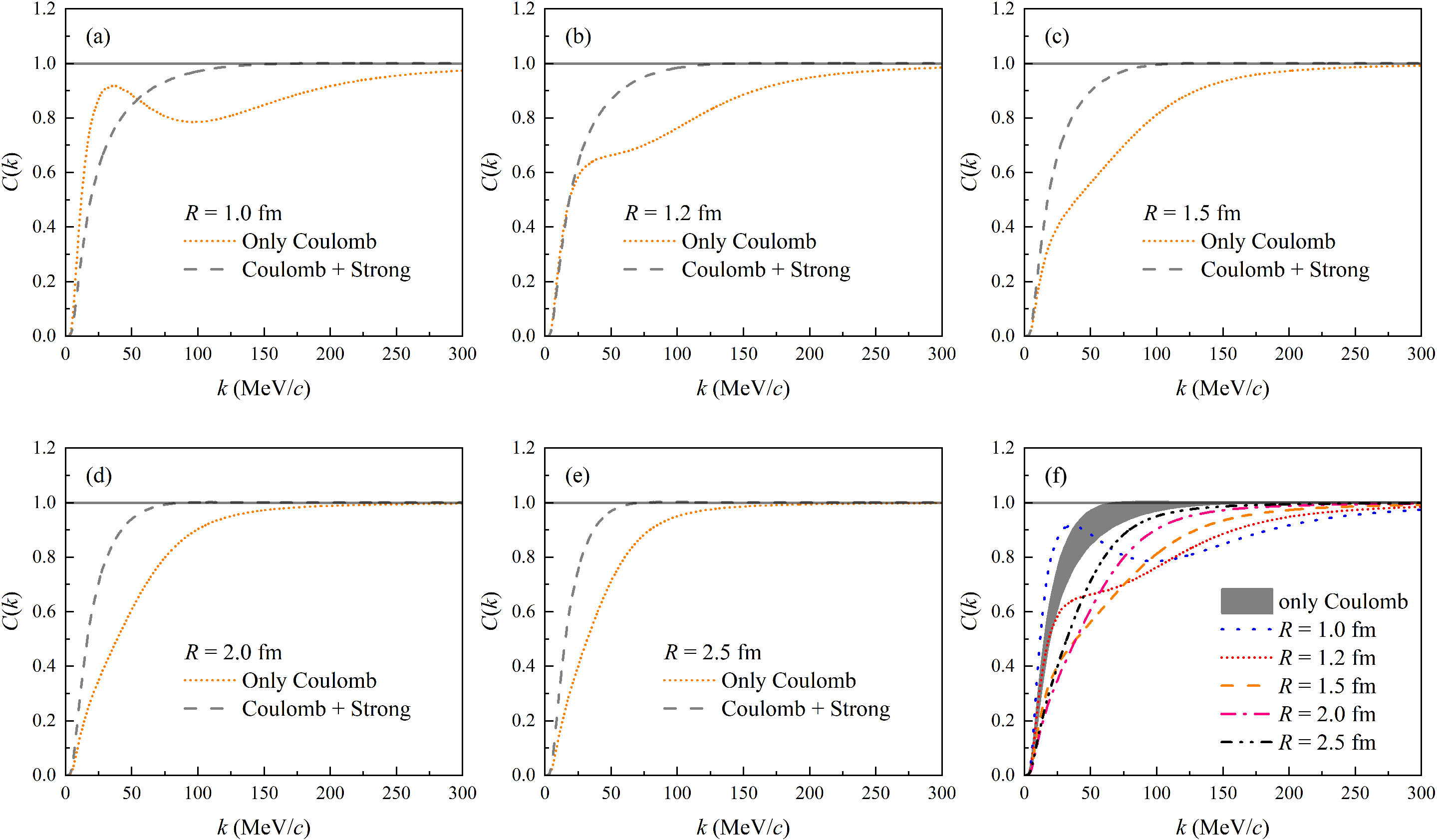}\
	\caption{The $p \bar{\Omega}$ correlation functions with different values of source size parameter $R$, where $R = 1.0$ fm in panel (a), $R = 1.2$ fm in panel (b), $R = 1.5$ fm in panel (c), $R = 2.0$ fm in panel (d), and $R = 2.5$ fm in panel (e) are summarized in panel (f).}
	\label{CF}
\end{figure*}

After replacing the square well potentials with the effective potentials obtained by solving the inverse scattering problem using the GLM method, which is briefly introduced in Section~\ref{GLM}, we can study the correlation function of the $p \bar{\Omega}$ systems.
Both the effective potentials of the $p \bar{\Omega}$ systems with $J^P = 1^-$ and $2^-$ are obtained.
The total $p \bar{\Omega}$ correlation function is the superposition of the correlation functions corresponding to the two quantum numbers according to Eq.~(\ref{average}).
Since the $p \bar{\Omega}$ system forms bound states for both quantum numbers and the interactions are similar, we omit the comparison of the correlation functions for the two quantum numbers here.
The total correlation functions calculated for different values of source size parameter $R$ are shown in Fig.~\ref{CF}.

In Fig.~\ref{CF}, panels (a)--(e), the dashed gray lines and the dotted orange lines represent the $p \bar{\Omega}$ correlation functions considering only the Coulomb interaction and both the Coulomb interaction and strong interaction, respectively.
In panel (f), the gray band stands for the correlation functions influenced only by the Coulomb interaction with size parameter $R$ ranging from 1.0 to 2.5 fm, while the other lines summarize the correlation functions shown in panels (a)--(e).
According to our results, the change of the size parameter $R$ can greatly influence the $p \bar{\Omega}$ correlation functions.
An obvious feature is that as $R$ increases, the peak-like structure caused by the different dominant regions of the Coulomb interaction and $p \bar{\Omega}$ strong interaction gradually becomes less obvious and eventually disappears.
Since two bound states are obtained in our calculation, it is very important to verify this conclusion in the correlation functions.
It can be seen that as the correlation function influenced only by the Coulomb interaction gradually approaches unity, the depletion caused by the bound states leads to the correlation function being below that of the Coulomb-only case.

In recent years, experimental data on correlation functions have grown rapidly, and with the help of femtoscopic technology, we can learn the interaction information of more systems.
At the same time, more theoretical work will help the development of this field.

\section{SUMMARY}

In this work, we investigate the $S$-wave $p \bar{\Omega}$ systems with isospin $I=1/2$, spin parity $J^{P}=1^{-}$ and $2^{-}$ in the framework of the QDCSM.
The results show that the $p \bar{\Omega}$ systems with both$J^{P}=1^{-}$ and $2^{+}$ form bound states, and the attraction between nucleon and $\bar{\Omega}$ is slightly larger than that between nucleon and $\Omega$, which indicates that the $p \bar{\Omega}$ is more possible to form bound state than the $p \Omega$ state.
The calculation of the low-energy scattering phase shifts and scattering parameters of the $p \bar{\Omega}$ systems also supports the existence of the $p \bar{\Omega}$ bound states with $J^{P}=1^{-}$ and $2^{-}$.
Besides, considering that nucleon is composed of three light quarks $u$($d$) and $\bar{\Omega}$ of three strange quarks $\bar{s}$, the $p \bar{\Omega}$ state cannot annihilate to the vacuum.
Therefore, the $p \bar{\Omega}$ state is a special state, which can provide useful information for the experimental search of the baryon-antibaryon bound states.

By using the GLM method, we solve the inverse scattering problem and obtain the effective $p \bar{\Omega}$ potentials.
On this basis, the $p \bar{\Omega}$ correlation functions are calculated and both the Coulomb interaction and spin-averaging effect are taken into account.
We present correlation functions corresponding to different source size parameters $R$, which can be used for future comparison with experimental measurements.

Understanding hadron-hadron interactions is one of the important issues in the study of quark models of hadron physics.
The study of the interaction between baryons and antibaryons in this work is also an effective place to test this mechanism.
The femtoscopic correlation function has become one of the important ways to explore hadron-hadron interactions, and more work is needed in the future.

\acknowledgments{This work is partly supported by the National Natural Science Foundation of China under contracts Nos. 11675080, 11775118, 12305087, 11535005, 11865019, 12135007, and 12233002. Q. W. is supported by the Natural Science Foundation of Jiangsu Province under grant No. BK20220122, the China Postdoctoral Science Foundation under grant No. 2024M751369, and the Jiangsu Funding Program for Excellent Postdoctoral Talent. Q. H. is supported by the Start-up Funds of Nanjing Normal University under grant No. 184080H201B20.}	

\setcounter{equation}{0}
\renewcommand\theequation{A\arabic{equation}}

\section*{Resonating group method for bound-state calculation and scattering process}

The resonating group method (RGM)~\cite{Wheeler:1937zza,Kamimura:1977okl} and generating coordinates method~\cite{Hill:1952jb,Griffin:1957zza} are used to carry out a dynamical calculation.
The main feature of the RGM for two-cluster systems is that it assumes that two clusters are frozen inside, and only considers the relative motion between the two clusters.
So the conventional ansatz for the two-cluster wave functions is
\begin{equation}
	\psi_{6q} = {\cal A }\left[[\phi_{B}\phi_{M}]^{[\sigma]IS}\otimes\chi(\boldsymbol{R})\right]^{J}, \label{5q}
\end{equation}
where the symbol ${\cal A }$ is the antisymmetrization operator, and ${\cal A} = 1 $ for the $p \bar{\Omega}$ systems.
$[\sigma]=[222]$ gives the total color symmetry and all other symbols have their usual meanings.
$\phi_{B}$ and $\phi_{M}$ are the $q^{3}$ and $\bar{q}q$ cluster wave functions, respectively.
From the variational principle, after variation with respect to the relative motion wave function $\chi(\boldsymbol{\mathbf{R}})=\sum_{L}\chi_{L}(\boldsymbol{\mathbf{R}})$, one obtains the RGM equation:
\begin{equation}
	\int H(\boldsymbol{\mathbf{R}},\boldsymbol{\mathbf{R'}})\chi(\boldsymbol{\mathbf{R'}})d\boldsymbol{\mathbf{R'}}=E\int N(\boldsymbol{\mathbf{R}},\boldsymbol{\mathbf{R'}})\chi(\boldsymbol{\mathbf{R'}})d\boldsymbol{\mathbf{R'}},  \label{RGM eq}
\end{equation}
where $H(\boldsymbol{\mathbf{R}},\boldsymbol{\mathbf{R'}})$ and $N(\boldsymbol{\mathbf{R}},\boldsymbol{\mathbf{R'}})$ are Hamiltonian and norm kernels.
By solving the RGM equation, we can get the energies $E$ and the wave functions.
In fact, it is not convenient to work with the RGM expressions.
Then, we expand the relative motion wave function $\chi(\boldsymbol{\mathbf{R}})$ by using a set of Gaussians with different centers
\begin{align}
	\chi(\boldsymbol{R}) =& \frac{1}{\sqrt{4 \pi}}\left(\frac{3}{2 \pi b^{2}}\right)^{\frac{3}{4}} \sum_{i,L,M} C_{i,L} \nonumber     \\
	&\cdot\int \exp \left[-\frac{3}{4 b^{2}}\left(\boldsymbol{R}-\boldsymbol{S}_{i}\right)^{2}\right] Y_{L,M}\left(\hat{\boldsymbol{S}}_{i}\right) d \Omega_{\boldsymbol{S}_{i}}
\end{align}
where $L$ is the orbital angular momentum between two clusters, and $\boldsymbol {S_{i}}$, $i=1,2,...,n$ are the generator coordinates, which are introduced to expand the relative motion wave function. By including the center-of-mass motion:
\begin{equation}
	\phi_{C} (\boldsymbol{R}_{C}) = (\frac{6}{\pi b^{2}})^{\frac{3}{4}} \, \text{exp}(-\frac{3}{b^2} \boldsymbol{R}^{2}_{C}),
\end{equation}
the ansatz Eq.~(\ref{5q}) can be rewritten as
\begin{align}
	\psi_{6 q} =& \mathcal{A} \sum_{i,L} C_{i,L} \int \frac{d \Omega_{\boldsymbol{S}_{i}}}{\sqrt{4 \pi}} \prod_{\alpha=1}^{3} \phi_{\alpha}\left(\boldsymbol{S}_{i}\right) \prod_{\beta=4}^{6} \phi_{\beta}\left(-\boldsymbol{S}_{i}\right) \nonumber \\
	& \cdot\left[\left[\chi_{I_{1} S_{1}}\left(B\right) \chi_{I_{2} S_{2}}\left(M\right)\right]^{I S} Y_{LM}\left(\hat{\boldsymbol{S}}_{i}\right)\right]^{J} \nonumber \\
	& \cdot\left[\chi_{c}\left(B\right) \chi_{c}\left(M\right)\right]^{[\sigma]}, \label{5q2}
\end{align}
where $\chi_{I_{1}S_{1}}$ and $\chi_{I_{2}S_{2}}$ are the product of the flavor and spin wave functions, and $\chi_{c}$ is the color wave function.
These will be shown in detail later.
$\phi_{\alpha}(\boldsymbol{S}_{i})$ and $\phi_{\beta}(-\boldsymbol{S}_{i})$ are the single-particle orbital wave functions with different reference centers,
\begin{align}
	\phi_{\alpha}\left(\boldsymbol{S}_{i}\right) & = \left(\frac{1}{\pi b^{2}}\right)^{\frac{3}{4}} \text{exp}[-\frac{1}{2 b^{2}}(\boldsymbol{r}_{\alpha}-\frac{1}{2} \boldsymbol{S}_{i})^2] , \nonumber \\
	\phi_{\beta}\left(-\boldsymbol{S}_{i}\right) & = \left(\frac{1}{\pi b^{2}}\right)^{\frac{3}{4}} \text{exp}[-\frac{1}{2 b^{2}}(\boldsymbol{r}_{\alpha} + \frac{1}{2} \boldsymbol{S}_{i})^2].
\end{align}
With the reformulated ansatz Eq.~(\ref{5q2}), the RGM Eq.~(\ref{RGM eq}) becomes an algebraic eigenvalue equation:
\begin{equation}
	\sum_{j} C_{j}H_{i,j}= E \sum_{j} C_{j}N_{i,j},
\end{equation}
where $H_{i,j}$ and $N_{i,j}$ are the Hamiltonian matrix elements and overlaps, respectively.
By solving the generalized eigenproblem, we can obtain the energy and the corresponding wave functions of the pentaquark systems.

For a scattering problem, the relative wave function is expanded as
\begin{align}
	\chi_{L}(\mathbf{R}) & =\sum_{i} C_{i,L} \frac{\tilde{u}_{L}\left(R, S_{i}\right)}{R} Y_{L,M}(\hat{\boldsymbol{R}}),
\end{align}
with
\begin{align}
	\tilde{u}_{L}\left(R, S_{i}\right) & = \left\{\begin{array}{ll}
		\alpha_{i} u_{L}\left(R, S_{i}\right), & R \leq R_{C} \\
		{\left[h_{L}^{-}(k, R)-s_{i} h_{L}^{+}(k, R)\right] R,} & R \geq R_{C}
	\end{array}\right.
\end{align}
where
\begin{align}
	u_{L}\left(R, S_{i}\right)= & \sqrt{4 \pi}\left(\frac{3}{2 \pi b^{2}}\right)^{\frac{3}{4}} R \,  \text{exp}[-\frac{3}{4 b^{2}}\left(R^{2} + S_{i}^{2}\right)] \nonumber \\
	& \cdot \text{i}^{L} j_{L}\left(-\text{i} \frac{3}{2 b^{2}} R S_{i}\right).
\end{align}

$h^{\pm}_L$ are the $L$th spherical Hankel functions, $k$ is the momentum of the relative motion with $k=\sqrt{2 \mu E_\text{{ie}}}$, $\mu$ is the reduced mass of two hadrons of the open channel, $E_{\text{ie}}$ is the incident energy of the relevant open channels, which can be written as $E_{\text{ie}} = E_\text{{total}} - E_\text{{th}}$, where $E_\text{{total}}$ denotes the total energy, and $E_\text{{th}}$ represents the threshold of the open channel.
$R_C$ is a cutoff radius beyond which all the strong interaction can be disregarded.
Additionally, $\alpha_i$ and $s_i$ are complex parameters that are determined by the smoothness condition at $R = R_C$ and $C_i$ satisfy $\sum_i C_i = 1$. After performing the variational procedure, a $L$th partial-wave equation for the scattering problem can be deduced as
\begin{align}
	\sum_j \mathcal{L}_{i j}^L C_j &= \mathcal{M}_i^L(i=0,1, \ldots, n-1),
\end{align}
with
\begin{align}
	\mathcal{L}_{i j}^L&=\mathcal{K}_{i j}^L-\mathcal{K}_{i 0}^L-\mathcal{K}_{0 j}^L+\mathcal{K}_{00}^L, \nonumber \\
	\mathcal{M}_i^L&=\mathcal{K}_{00}^L-\mathcal{K}_{i 0}^L,
\end{align}
and
\begin{align}
	\mathcal{K}_{i j}^L= & \left\langle\hat{\phi}_A \hat{\phi}_B \frac{\tilde{u}_L\left(R^{\prime}, S_i\right)}{R^{\prime}} Y_{L,M}\left(\hat{\boldsymbol{R}^{\prime}}\right)|H-E|\right. \nonumber \\
	& \left.\cdot \mathcal{A}\left[\hat{\phi}_A \hat{\phi}_B \frac{\tilde{u}_L\left(R, S_j\right)}{R} Y_{L,M}(\hat{\boldsymbol{R}})\right]\right\rangle .
\end{align}
By solving Eq.~(A11), we can obtain the expansion coefficients $C_i$, then the $S$-matrix element $S_L$ and the phase shifts $\delta_L$ are given by
\begin{align}
	S_L&=e^{2 i \delta_L}=\sum_{i} C_i s_i.
\end{align}

Resonances are unstable particles usually observed as bell-shaped structures in scattering cross sections of their open channels.
For a simple narrow resonance, its fundamental properties correspond to the visible cross section features: mass $M$ is at the peak position, and decay width $\Gamma$ is the half-width of the bell shape.
The cross section $\sigma_{L}$ and the scattering phase shifts $\delta_{L}$ have relations
\begin{align}
	\sigma_L&=\frac{4 \pi}{k^2}(2 L+1) \sin ^2 \delta_L.
\end{align}
Therefore, resonances can also usually be observed in the scattering phase shift, where the phase shift of the scattering channels rises through $\pi/2$ at a resonance mass.
We can obtain a resonance mass at the position of the phase shift of $\pi/2$.
The decay width is the mass difference between the phase shift of $3\pi/4$ and $\pi/4$.

\end{document}